\documentclass[letterpaper,twocolumn,10pt]{article}
\usepackage{usenix2022_SOUPS}

\usepackage{tikz}
\usepackage{amsmath}
\usepackage{graphicx}
\usepackage{textcomp}
\usepackage{tabularx}
\newcolumntype{Y}{>{\centering\arraybackslash}X}

\newcolumntype{Z}{>{\hsize=1.3\hsize}X}
\newcolumntype{Q}{>{\hsize=.7\hsize}Y}
\newcolumntype{V}{>{\hsize=.15\hsize}X}
\usepackage{caption}
\usepackage{subcaption}
\usepackage{multirow}
\usepackage{multicol}
\usepackage{makecell}
\usepackage{xcolor}
\usepackage{xurl}
\usepackage{hyperref}
\usepackage{enumitem}
\usepackage{censor}
\usepackage{booktabs}

\usepackage{filecontents}

\begin{document}

\date{}

\title{\Large \bf Fight Fire with Fire: Hacktivists' Take on Social Media Misinformation}

\def\plainauthor{Author name(s) for PDF metadata. Don't forget to anonymize for submission!}

\author{
{\rm Filipo Sharevski}\\
DePaul University
\and
{\rm Benjamin Kessell}\\
DePaul University
} 

\maketitle


\begin{abstract}
In this study, we interviewed 22 prominent hacktivists to learn their take on the increased proliferation of misinformation on social media. We found that none of them welcomes the nefarious appropriation of trolling and memes for the purpose of political (counter)argumentation and dissemination of propaganda. True to the original \textit{hacker} ethos, misinformation is seen as a threat to the democratic vision of the Internet, and as such, it must be confronted on the face with tried hacktivists' methods like deplatforming the ``misinformers'' and doxing or leaking data about their funding and recruitment. The majority of the hacktivists also recommended interventions for raising misinformation literacy in addition to targeted hacking campaigns. We discuss the implications of these findings relative to the emergent recasting of hacktivism in defense of a constructive and factual social media discourse.
\end{abstract}

\section{Introduction}
Steven Levy's portrayal of the hacker culture in his 1984 book \textit{Hackers}  largely remains the most influential reference to the public's general view of hackers \cite{Levy, Hatta}. Recasting them Robin Hood-style activists committed to a democratic vision of the Internet \cite{Rolfe}, Levy asserts that the hacker ethos embodies several sacrosanct postulates to the public good, notably that (i) \textit{all information should be free}, and (ii) \textit{authority should be mistrusted and decentralization promoted} \cite{Levy}. 

Later-day Internet hackers shifted the ideological tendency for autonomy in the cyberspace towards a vision of the Internet as a popular space for sharing any information that can nevertheless be politicized and weaponized against the neoliberal elites responsible for economic and social disarray \cite{Gerbaudo}. Turning Internet activism into a form of socio-political resistance online \cite{Karagiannopoulos}, enabled a functional selection of issues that no longer necessitated a long preparation \cite{McCaughey}. This, in turn, resulted in almost instant convergence and coordination of activities in response to the issues of interest that, over the years, became publicly visible through mass media coverage \cite{Illia}. 

The Internet activism, expectedly, bifurcated to online campaigns concerned with the protection of the Internet as a relatively unregulated and unowned space (e.g. Anonymous,  WikiLeaks, Snowden \cite{Coleman, Sifry, Snowden}) and online campaigns concerned with the protection of human rights and the environment (e.g. the Occupy movement, Arab Spring, Pirate Party \cite{Milan, Karatzogianni}). The former activism -- or \textit{hacktivism} - often is anonymous, performed in secret, and operates with a kind of impunity that the Internet technologies seem to afford so far \cite{Sorell}. The later activism -- or \textit{hashtag activism} -- usually is public, openly used the Internet for political mobilization, operates primarily on the streets, and subjects to the dangers of crowd violence, harassment, and arbitrary arrest  \cite{Sandoval-Almazan}. 

The hashtag activism historically utilized various Internet technologies such a petition websites (e.g. MoveOn.org for organizing political protests) or e-mail communication (e.g. Tea Party’s campaign to reduce government spending and taxation) \cite{Carty}, but the advent of social media sites like Twitter, Facebook, and YouTube truly accelerated the self-organization and participation in the sociopolitical struggle (e.g. the \#BlackLivesMatter and \#SchoolStrike4Climate movements \cite{Freelon}). While the essential dependence on social media is apparent, both in a historical context and for the future of the hashtag activism \cite{Jungherr}, the relationship between the hacktivism and social media is a bit more complicated. 

Hacktivists, in contrast, hacked various Internet technologies such as defacing websites \cite{Romagna}, breaking into systems to ``leak'' and ``dox'' private documents \cite{Sifry, Steinmetz}, and storm systems with traffic to cause a Denial-of-Service (DOS) \cite{Milan-hacktivism}. Hacktivists' foray in social media mirrors these actions as campaigns were undertaken for hijacking/defacement of social media accounts (e.g., Anonymous's \#OpKKK campaign \cite{Wright}), doxing individuals on Twitter (e.g. the students of Covington High School \cite{Lindvall}), and DoS Twitter topics (e.g. \#IranTalks campaign \cite{Najafabadi}). But hacktivists also hacked the social media affordances for content amplification (e.g. StayWokeBot \cite{George, Savage}), early instances of trolling (e.g. Rickrolls \cite{Sanfilippo}), and sharing memes (e.g. Lol Cats on 4chan \cite{Coleman}). 

Despite the intuitive versatility of social media for such subversive operations,  hacktivism became largely inactive on the mainstream platforms following some high profile run-ins with the legal authorities of the leading hacktivists \cite{Jones, Uitermark}. The apparent absence of hacktivism created a vacuum where no one actively challenged the elites, defended freedom of expression, and appended the vision of democratic social media participation. It took little time, unfortunately, for this vacuum to be appropriated by state-sponsored actors hijacking the hacking playbook for actions aimed not just against the neoliberal elites but the entire social order \cite{Follis}. Bot-style amplification aided political trolling and sharing of memes in the beforemath of Brexit campaign in the UK \cite{Corner} and the 2016 elections in the US \cite{Benkler}. The crucial difference in these instances was that the amplified memes and trolling were not pranks but damaging fake news, emotionally-charged memes, and conspiracy theories that instead of unifying the social media crowds for a cause, divided them in opposition camps pitted against each other \cite{VoxPop}.     

In response to such a large-scale disruption on the social media turf, one would have plausibly expected that the hacktivists will retaliate and confront, expose, or counter hack the state-sponsored ``trolls'' \cite{Caulfield}. Misinformation, back to the Levy's depiction of hacker's ethics \cite{Levy}, runs counter the  
(i) \textit{all information should be free} postulate because it undermines the basic utility of information as a public good (i.e. truth and facts do not dwindle in supply as more people ``consume'' them and truth and facts are available to all people in a society) \cite{Ehsanfar}. Misinformation also runs counter the (ii) \textit{authority should be mistrusted and decentralization promoted} postulate because it is promulgated by a state-sponsored ``shadow authority,'' as evidence confirms in the aftermath of the Brexit and the 2016 US elections \cite{Llewellyn, Zannettou2020, Im}. Surprisingly, the hacktivists never struck back \cite{Beraldo}, though they clearly poses the capabilities to do so, as witnessed in the Anonymous's \#OpISIS campaign, for instance, where the collective flagged about 101,000 Twitter accounts attributed to the Islamic-State \cite{Ireland}. 

The absence of response to misinformation on social media by the hacktivist community seemed quite perplexing and, in our opinion, worthy of in-depth inquiry with active ``hackers'' that still operate in the spirit of the Levy's code of ethics \cite{Levy}. Through personal connections and snowballing sampling, we identified 22 prominent hacker figures and set down for at least an hour-long interview with each of them to learn their take on the misinformation ecosystem, on responses to falsehoods on social media, and the way misinformation impacts and shapes the hacktivists' agenda in the future. We found a consensus among the hacktivists against the present forms of misinformation as an ammunition for political counter(argumentation) and external propaganda. They were adamant to deplatform, dox, and expose every ``misinformer'' that they believe is polluting the social media discourse, and suggested ways to improve the general misinformation literacy among users in addition to these targeted operations. 

To situate our study in the intersection between the hacktivist counter-culture and the rise of misinformation on platforms, we review the interplay between Internet activism, social media, and false information in Section \ref{sec:litreview}. We look in the broader context of misinformation in Section \ref{sec:m/disinfo} to highlight the pressing need of hacking action to reclaim the social media space true to Levy’s vision of Internet as an information exchange to the public good. In Section \ref{sec:study} we outline our research design and methodology. Sections \ref{sec:results-rq1}, \ref{sec:results-rq2}, and \ref{sec:results-rq3} expand on our findings and we discuss the implications of the hackers' disposition to social media misinformation in Section \ref{sec:discussion}. Finally, Section \ref{sec:conclusion} concludes the paper.

\section{Internet Activism and Social Media} \label{sec:litreview}
\subsection{Hashtag Activism}
Online social media activism -- or \textit{slacktivism}, \textit{clicktivism}  -- emerged on popular platforms as a repertoire of low-risk, low-cost expressive activities for advocacy groups' agenda setting and political participation \cite{Rotman}. Social media users participated in petitions, changed personal avatars, added picture filters in support of a cause, and simply ``liked'' posts as an act of participation \cite{Halupka}. Slacktivists quickly realized they could use virality as a distinctive social media affordance to their advantage and move to use hashtags as the main drivers of mobilization, raising awareness, and demanding sociopolitical change. The practice of \textit{hashtag activism} was instrumental for the success of social movements like \#metoo, \#takeaknee, and \#BlackLivesMatter, allowing for mainstream visibility, expression of solidarity, and statement of victimhood \cite{Simpson}. This success, in turn, inspired a plethora of other movements advocating for health, human rights, social justice, and environmental issues to spur across all social media platforms and remain active within the public discourse \cite{Jackson}. 

The materialization of the hashtag activism, however noble, had to deal with the obvious threat of \textit{hashtag hijacking} or the encroachment on viral hashtags to inject contrary perspectives into a discourse stream \cite{VanDam}. This ``hack'' against the internet activism is not just adding noise or attempting to result in a DoS, but also to disseminate hateful narratives and dilute the campaign itself (e.g. the hijacking of the \#metoo hashtag \cite{Lindgren}). Another similar threat is the \textit{hashtag co-opting} or the contentious co-opting of the rhetoric of popular social movements (e.g. \#HeterosecualPrideDay campaign co-opting the language of the mainstream LGBT movement \cite{Armstrong}). Equally threatening is the \textit{counter hashtagging} that concocts similar hashtags to garner opposition to well-established movements (e.g. \#BlueLivesMatter countermovement to police reform in reaction to \#BlackLivesMatter \cite{Vance}). These antagonistic appropriations of the social media virality, consequently, enabled political extremism to creep in the public discourse and embroil users in an emotionally-charged participation \cite{Rho}. 

In a state of emerging social media polarization, it was a question of time when fake news, offensive memes, and conspiracy theories would be weaponized against the hashtag activism (e.g. the proliferation of fake news in the \#Gunreformnow vs \#NRA Twitter battle \cite{Chong}). What was initially expected to remain on the fringes of the mainstream hashtag activism \cite{Freelon-Marwick}, quickly turned into an information disorder on a mass scale. Now the hashtag hijacking and co-opting developed \textit{in parallel} with the main theme of activism, and for that, a steady and substantive feed of false and unverified information was needed. The emotionally-charged participation loomed into a global health panic (e.g. \#FlattenTheCurve hashtag hijacking for spreading COVID-19 misinformation \cite{Philipp}) and moral panic (e.g. the QAnon's co-opting of \#SaveTheChildren hashtag \cite{Moran}) in addition to the already growing political panic \cite{Mustafaraj}.

\subsection{Hacktivism}
\textit{Hacktivism} was a term that ``Omega,'' a member of the Texas-based computer-hacking group \textit{Cult of the Dead Cow} (cDc) coined in 1996 in an email to the cDc listserv \cite{McCormick}. Characterized with the increasingly political ethos of hacking-for-cause, hacktivists primarily leveraged technology to advance human rights and protect the free flow of information in campaigns against the UK, US, and Chinese governments, as well as the UN \cite{Owen}. In as much as hackers individually roamed the Internet, socialization was at the proper time as many of them needed establishing a strong hacktivist network. Hacktivists' penchant for humorous memes (LOLCats) and gag hyperlinks (Rickrolls) \cite{Phillips} attracted an army of hackers to Christopher Poole's \text{4chan.org} social media website, bringing to life the notorious collective Anonymous \cite{McCormick}.

While hacktivists never displayed a predictable trajectory of their cyberoperations and political program \cite{Coleman}, they narrowly utilized social media for self-promotion --  announcing operations with an \#Op prefixed hashtags \cite{Beraldo} -- and furthering a complex relationship with other Internet activists. Anonymous cried foul on Twitter when WikiLeaks puts millions of its documents behind a pay wall \cite{Hallsby}, but also launched Operation \#Ferguson of doxing the St. Louis County police chief daughter's information in response to the shooting of the black teenager Michael Brown \cite{Bellaby}. Hacktivists, in solidarity to the Arab spring uprisings, sent a care package composed of security tools and tactical advice though downplayed the touted ``Twitter Revolution'' \cite{Coleman}.

True to their credo for utilizing Internet technologies against oppression, including social media, hacktivists launched the \#OpKKK to ``unhood approximately 1000 Ku Klux Klan members'' hacked by gaining access to a KKK Twitter account in support of \#BlackLivesMatter protesters in Ferguson, Missouri \cite{Wright}. After a several years hiatus, perhaps due to arrests of some of the leading Anonymous hacktivists, the group resurfaced during the 2020 \#BlackLivesMatter protests in response to the killing of George Floyd \cite{Jones-Nurse}. This time, in addition to the leaking of a trove of 269 gigabytes of confidential police data (dubbed \textit{BlueLeaks} \cite{Lee}), the hacktivists launched social bot operations to amplify a support towards \#BLM and criticize police actions.

Hacktivists also utilized Internet technologies in the context of cyberwarfare. For example, the \#OpIsis operation, in which lists of tens of thousands of Twitter accounts that purportedly belonged to members of ISIS or its sympathizers were leaked, was launched in response to the terrorist attacks in France in 2015 \cite{McGovern}. Here, in addition to the leaks, hacktivists also waged a meme war and called for a ``Troll ISIS Day'' to provoke and disrupt ISIS-supported social media \cite{McCrow-Young}. The Anonymous group in early 2022 took on Twitter to declare a ``cyber war'' to Russia in response to the Ukrainian invasion, launching DoS attacks against Russian's Federal Security Service’s website and hacking Russian streaming services to broadcast war videos from Ukraine \cite{Serpanos}.

\section{Internet Activism and Misinformation} \label{sec:m/disinfo}
\subsection{Grassroots Misinformation Operations}
Hacktivists, perhaps inadvertently, authored or gave popularity to the most utilized primitives for creating, propagating, amplifying, and disseminating misinformation - \textit{trolling} and \textit{memes}. This negative externality is unfortunate as trolling and memes were initially used by Anonymous against what they perceived a ``misinformation campaign'' by the Church of Scientology  \cite{McCormick}. The ``anon'' members on \texttt{4chan.org} practically \textit{hijacked} the term ``troll'' -- initially meaning provoking others for mutual enjoyment -- to abusing others for members' own enjoyment by posting upsetting or shocking content (usually on the \textbackslash{b} channel of \texttt{4chan.org} \cite{Coleman}), harassing users (e.g. mocking funeral websites \cite{Bishop}), and spreading rumors \cite{Klempka}. What Anonymous did for the ``lulz'' (a brand of enjoyment etymologically derived from laughing-out-loud (lol)), nonetheless, showed the ease with which one could exploit the Internet technologies to be impolite, aggressive, disruptive, and manipulative to users' emotional states \cite{Coleman}. 

Trolling initially came in textual format as comments to posts, bulletin boards, and websites ``deindividualized'' people's lived experience for the ``lulz'' \cite{Bishop}. Gradually, hacktivists popularized a multimedia format of trolling or ``memes,'' where textual commentary is superimposed over well-known imagery, typically representing different forms of power, such as political leaders, the police, and celebrities \cite{McCrow-Young}. Memes, perhaps, were the actual rite of passage to true hacktivism -- moving away from the early LOLCats -- as they seek  to deconstruct the power represented, contest censorship, and provide political commentary \cite{Nissenbaum}. Memes as content were put to hacktivist use \textit{en masse} in operations like ``Troll ISIS day,'' where Anonymous proliferated memes with rubber-duck heads or rainbow stripes to ridicule ISIS propaganda imagery and disinformation narratives on Twitter \cite{McCrow-Young}. Spread together with satirizing hashtags (e.g. \#Daeshbags), the trolling memes achieved a cultural virality that brought hacktivists' into the mainstream discourse online \cite{Phillips-Fox}. What the hacktivists did with the memes nonetheless, showed the ease with which one could disrupt, challenge, reimagine, and appropriate new political contexts by harnessing the virality and visibility of content spread on social media \cite{Mortensen}.

\subsection{Mainstream Misinformation Operations}
The hacktivists' playbook of trolling and meme dissent, though initially targeted \textit{against} misinformation, was skillfully appropriated \textit{for} crafting and disseminating misinformation from 2014 onward, coinciding with the period of   hacktivist inactivity \cite{Beraldo}. The playbook alone, at first, was insufficient to the objectives of widespread political disruption as it necessitated a support network of individuals and/or accounts on social media for any alternative narratives to gain traction. But the ``appropriators'' -- privy of prior campaigns of disinformation and with the support of nation-state governments \cite{Sharevski-Facebook} -- need not to look further as ``sock puppet'' accounts were already utilized for spreading political falsehoods (e.g., Martha Coackey's ``twitter bomb'' disinformation campaign \cite{Mustafaraj}). Having all the ingredients for exploiting the virality of social media and users' familiarity with emotionally-charged discourse, the ``appropriators'' established \textit{troll farms} in the wake of the UK's Brexit campaign and 2016 US elections \cite{Llewellyn, Caulfield}. 

The ``army'' behind the troll farms were particularly clever to append their social bots with ``sock puppet'' accounts that imitate ordinary users to systematically micro-target different audiences, foster antagonism, and undermine trust in information intermediaries \cite{Arif}. Playing both sides in the emotionally-charged discourse already unravelling on social media, the troll farms posed as authentic, cultural competent personas (e.g. so-called ``Jenna Abrams'' account \cite{Xia}), as well as vocal supporters of hashtag activism (counter) movements (e.g. \@BlackToLive in \#BlackLivesMatter and \@SouthLoneStar in \#BlueLivesMatter \cite{Stewart}). They also appropriated the hashtag hijacking (e.g., \#elections2016 and \#ImVotingBecause tagging of quotes about Donald Trump and against Hillary Clinton \cite{Alonso}), hashtag co-opting (e.g. \#BlackGunsMatter and \#syrianlivesmatter \cite{DiResta}), and counter hash tagging (e.g. \#NoDAPL against the Dakota Access Pipeline \cite{Hindman}). The troll farms even had the audacity to impersonate the Anonymous themselves (e.g. the @\_anonymous\_news impersonation of the ``Your Anonymous News'' twitter account \cite{Coleman2020}). 

The ``meme game'' of the troll farms was equally sophisticated and appended the initial success of their operations \cite{Milner}. Testing the waters with war-related memes regarding the opposition/support of the conflict in Syria \cite{DiResta}, the troll farms capitalized on both the meme trolling and the Internet activism by spreading political memes through their \@Blacktivist social media accounts and co-opting Wikileaks in exploiting the leak of sensitive documents from the Democratic National Committee (DNC) \cite{Linvill}. Memes were also used to amplify conspiracy theories (e.g. QAnon, Pizzagate, and the murder of Seth Rich \cite{ZannettouGab}), Texas secessionism (e.g. if Brexit why not \#Texit \cite{TrollHunter2020}), and direct attacks (e.g. crooked Hillary \cite{Howard}).

While the initial campaigns of the troll farms have been tracked, exposed, and brought into attention \cite{DiResta, Howard}, the social media discourse never really recovered from the watershed appropriation of the Internet activism for the purpose of conducting information operations \cite{VoxPop}. Worse, the troll farm brand of political dissent was adopted by populist accounts that were keen on disseminating misinformation beyond just politics \cite{TrollHunter}. The trolling pandemonium spilled out of control with the COVID-19 pandemic as rumors, conspiracy theories, fake news, and out-of-context spins plagued the social media by hijacking the dominant hashtags like \#COVID19, \#coronavirus or \#DoctorsSpeakUp \cite{Bradshaw}, co-opting hashtags like \#plandemic \cite{Kearney} and counter hash tagging with hashtags like \#COVIDIOT \cite{Sharevski-Huff}. Memes were distributed in conjunction with deepfake videos on platforms like YouTube \cite{Rochert} and TikTok \cite{Basch} as well as blatant fake news on alt-platforms like Gab \cite{Cinelli} to effectively reach a self-perpetuating bedlam of misinformation Internet counter-activism.

\section{Hacktivism and Misinformation} \label{sec:study}
In a radical state of ravaging misinformation campaigns on social media with no end in sight, one could wonder what the original activists on the Internet have to say in response. The unravelling of falsehoods clearly is a serious threat to the democratic vision of the Internet \cite{Rolfe}, as misinformation facilitated the rise of non-democratic communities contesting even factual knowledge and science (e.g. anti-vaxers, climate change deniers, etc. \cite{Waisbord}). Hacktivists, as we have seen in Section \ref{sec:litreview}, have fiercely opposed early misinformation campaigns in the past, but their means to do so were the ``hijacked'' for the mass produced misinformation of later. One could attribute the paucity of hacktivists' involvement in the passing of the techno liberal order of the Internet as the rise of partisan-divided trust in facts and the politicization of science were already underway \cite{Gauchat}, but that alone is not a sufficient showstopper for action.

Regardless of any new Internet order, there is a reasonable expectation that one should still act upon the Levy's sacrosanct postulates \cite{Levy}, even if that is within an ecosystem polluted with misinformation. In addition to the public good arguments, misinformation is in conflict with the \textit{all information should be free} postulate as it creates ``information disorder'' that, by the token of catalyzing polarization and emotionally-charged participation online, gives even more power to the neoliberal elites for perpetuating the economic and social (media) disarray \cite{Creech}. Misinformation also conflicts with the \textit{authority should be mistrusted, and decentralization promoted} postulates as it stands in the way of independent truth discovery and dissemination online \cite{Lewandowsky}. Should the new brand of reprehensible misinformation, therefore, be on the top of the hacktivists' agenda already? 

\subsection{Research Questions}
To explore the gap in hacktivism in regards mass misinformation, we invited prominent members of the hactivist community to answer the following research questions:  

\begin{itemize}
    \item \textbf{RQ1:} How contemporary hacktivists conceptualize the social media misinformation ecosystem?  
    \item \textbf{RQ2:} What action hacktivists deem appropriate in responding to misinformation on social media? 
    \item \textbf{RQ3:} In what directions do the hacktivists see the misinformation ecosystem evolve in the future?
\end{itemize}

\subsection{Sample}
Our study was approved by the Institutional Review Board (IRB) of our institution before we invited, through personal contacts, and snowballing sampling the hacktivists for a virtual interview session with open-ended questions, listed in the \hyperref[appendix]{Appendix}. We sampled a population who were 18 years or older, from the United States, that is an active contributor in the hacktivist community, 
and has a history of such an involvement that we could reasonably verify. We used zoom interviews where we offered the possibility for the participants to choose if they want to use a video feed or not. Every interview was recorded, stored in a secure server, and manually transcribed and communicated with the interviewee to obtain an approval before we started the qualitative analysis. 

Overall, we ended with a sample of total of 22 participants, all of which agreed to participate voluntarily. The demographics are given in Table \ref{tab:demographics}.  We made a deliberate attempt to produce a sample that is not a male-only or male-dominated, as previous studies indicate that the hacktivist community is imbalanced in regards gender \cite{Tanczer}. The participation in the study was not anonymous to us as researchers, but we deliberately avoid using definitive numbers and potentially identifiable information in reporting of our results to preserve their anonymity to the general population, as a condition for their participation. In some cases, we used a direct censoring of names in citing participants' responses. We allowed the participants to skip any question they were uncomfortable answering. The interviews took around an hour to complete. 

\begin{table}[htbp]
\renewcommand{\arraystretch}{1.3}
\footnotesize
\caption{Sample Demographic Distribution}
\label{tab:demographics}
\centering
\aboverulesep=0ex 
   \belowrulesep=0ex 
\begin{tabularx}{\linewidth}{|Y|}
\Xhline{3\arrayrulewidth}
\toprule
 \textbf{Gender} \\\Xhline{3\arrayrulewidth}
\midrule
\footnotesize
\vspace{0.2em}
    \hfill \makecell{\textbf{Female} \\ 8 (36.4\%)} 
    \hfill \makecell{\textbf{Male} \\ 13 (59.1\%)} 
    \hfill \makecell{\textbf{Non-Binary} \\ 1 (4.5\%)} \hfill\null
\vspace{0.2em}
\\\Xhline{3\arrayrulewidth}
\end{tabularx}
\end{table}

\subsection{Methods and Instrumentation}
To ensure validity to the task of conceptualizing misinformation, we decided to introduce the participants in the main study to the generalized definition of misinformation on social media proposed by Wu et al. \cite{Wu}. Another reason was to avoid confusion between past trolling and memes ``for the lulz'' and present alternative narratives that involve information operations, rumors, conspiracy theories, fake news, hoaxes, and clickbait. The hacktivists in our sample were invited to speak about their profiles, activity, and agendas online, before we asked their take on misinformation on social media. The qualitative responses were coded and categorized in respect: a) antecedents to misinformation; b) mental models of misinformation; c) countering misinformation through leaking, doxing, and deplatforming; d) anti-misinformation ``ops'' (operations); e) counter-misinformation tactics; f) misinformation literacy; and g) misinformation hacktivism. 

Two independent researchers analyzed the approved interview transcriptions, achieving a strong level of inter-coder agreement (Cohen's $\kappa = .82$). We utilized a thematic analysis methodology to identify the themes and sub-themes most saliently emerging from the responses in our sample. The themes were summarized to describe the conceptualization, response, and evolution of misinformation in the view of the contemporary hacktivists we sampled. In reporting the results, we utilized as much as possible verbatim quotation of participants' answers, emphasized in ``\textit{italics}'' and with a reference to the participant as either \textbf{PX} or [\textbf{PX}], where \textbf{P} denotes \textbf{participant} and \textbf{X} denotes the \textbf{number} of the participant in the sample (ordered by the time of participation).

\subsection{Hacktivists' Profiles}
The hacktivists in our sample, true to the original ethos, represent the voice for advocacy and contemporary policy discussion. While they did not disclose their current operations, several of them hinted they are involved in tracking the rise of the far-right extremism, cybercriminals, as well as the information warfare part of the Ukraine invasion. A couple of the hacktivists' agenda was leaking documents from companies and nation-state agencies as manifestation of their information freedom advocacy. Few of the hacktivists explicitly mentioned they still create and disseminate memes and participate in the ``old school'' trolling. And several of the hacktivists did actual \textit{hacking} as in analyzing security problems (e.g. ransomware) and providing free tools for helping ordinary Internet users fend off related threats. 

The majority of the hacktivists noted they have been active for a long time, being brought into the world of computers in childhood or early adolescence. Some of them resorted to hacktivism  as  a way to protect themselves against online bullies and some of them in response to nation-states offensive operations online, notably ones linked to China and Russia. Several of them started with hacking operating systems to enable unrestricted access to games and/or bypass parental controls. While most of the participants in our sample cited curiosity as their driver to enter the ``hacktivist conglomerate'' and keep on hacking, there were some participants citing a deliberate determination for cybersecurity education activism.

\section{Misinformation Conceptualization} \label{sec:results-rq1}
Social media users conceptualize misinformation, evidence shows, in more than one model that narrowly focuses on inherently fallacious information \cite{folk-models}. Beyond just fake news, misinformation is equally conceptualized as form of \textit{political (counter)argumentation} where facts do selectively appear in alternative narratives relative to political and ideological contexts, often taken \textit{out-of-context} with speculative intentions. Misinformation is also seen as \textit{external propaganda} that includes \textit{manufactured} facts and factoids disseminated and amplified online with division-creating intentions. Given the radical transformation of the trolling and mimes over time, our first research question aimed to learn the hacktivists' take on this transformation in the context of the competing conceptualizations amongst ordinary social media users. 

\subsection{Antecedents to Misinformation}
The participants in our sample agree that trolling and mime dissemination has been hijacked for nefarious purposes, lamenting that what was a ``\textit{deliberate action mostly for laughs, now is an automated operation for keeping people tribalistic
and resistant to opposing views}'' [\textbf{P13}]. The use of ``\textit{sock puppets for running forum raids in the old days of hacktivism}'' [\textbf{P4}], unfortunately, was not enough a serious threat for social media to implement ``\textit{strict policies of who and how can participate in the public discourses early on}'' [\textbf{P1}] and  counter to their business model of ``\textit{monetizing every possible engagement on their platforms} [\textbf{P14}].

Mainstream social media companies were accused of directly enabling the ``information disorder'' as their models of engagement pushed ``\textit{less educational content the more an issue was important and demanded action}'' [\textbf{P14}]. This disorder played in the hands of the neoliberal elites, media outlets, and news organizations run by ``\textit{billionaires detached from reality to gain further control over public spaces}'' as \textbf{P1} put it. In the view of our participants, misinformation ``\textit{has always been there}'' and pointed to the combination of ``\textit{self-proclamation of expertise online, cultivating followers, and playing on confirmation bias}'' as the recipe the very hacktivists showed it works well in seeding misinformation: 

\begin{quotation}
``\textit{For example, look at the \censor{H4unt3d Hacker}. He said he was a founding member of Anonymous and lots of people believed him. He has spoken at conferences about it and even got jobs because of it. Literally dig slightly into that and it's clear that no one in the Anonymous community can vouch for the guy and there's no evidence of him being linked. So, people are just too lazy to check stuff out because this guy is kinda selling a story that fits with what they think so it must be true}'' [\textbf{P3}]. 
\end{quotation}

\subsection{Mental Models of Misinformation}
The predominant mental model of misinformation amongst the hacktivists in our sample was the \textit{political (counter)argumentation} where the information disseminated on social media for the sake of furthering a political argument or agenda \cite{folk-models}. In the original version of trolling and meme sharing the misinformation was seen as an alternative expression of disagreement, revolt, or ridicule without any context, but the contemporary trolling and memes is brought in the political context as a ready-made content for expression of political attitudes \cite{Pennycook-Rand-Psych}. Despite that fact checking is widely available (and even suggested to users when content is moderated on social media \cite{context2022}), the political appropriation of misinformation thrives because ``\textit{people won't fact check things and perpetuate them as long as these things align with their political ideology}'' [\textbf{P2}]. The reason why most social media users ``\textit{fall for misinformation}'' is \textit{plain ignorance and stubbornness to hear anything contrary to their own political opinions}'' [\textbf{P3}] which results from ``\textit{a serious lack of, at least in the U.S, critical thinking education in schools} [\textbf{P2}]. 

In the view of the majority hacktivists in our study,``\textit{both sides of the political spectrum spread misinformation and it further enables political polarization}'' [\textbf{P13}]. While they acknowledge that ``\textit{the misinformation on social media is often identified with right-wing opinions}'' [\textbf{P6}], hacktivists recognize that ``\textit{we overuse the terms misinformation and disinformation to describe anything that is not a leftist opinion or fact} [\textbf{P7}]. They point to the misinformation ``stickiness'' where the repeated exposure to speculative and false statements make them appear truthful \cite{debunking-handbook}, becoming the main theme of every social media discourse. For example, \textbf{P3} refers to the Biden's laptop saga \cite{Higdon}, which in their view ``\textit{has been politically disinfoed [sic] to death to the point that the laptop leaks are irrelevant and can't be trusted as an evidence}.'' 

Misinformation as political counter(argumentation) bothers the hacktivists as it conflicts with the \textit{all information should be free} postulate, which in turn forces mainstream social media platforms to ``\textit{restrict the flow of information}'' [\textbf{P10}]. Misinformation, in the view of \textbf{P10}, should not be restricted because ``\textit{people are entitled to see both sides of a proverbial political coin so the platforms must allow them to do so, otherwise by only showing heads or tails people will speculate about what's on the other side and assume the worst}.'' The restriction of information on platforms conflicts with the \textit{mistrust of authority and promote decentralization} hacker postulate because ``\textit{self-appoints the elites to define what constitutes `truth'}'' [\textbf{P14}]. It also forces ``\textit{people to become rather tribalistic and a priori suspicious of people with different views}'' [\textbf{P}]. The ``political tribalism'' on social media \cite{Allen}, in turn, makes it ``\textit{easier to demonize people with different opinions and political attitudes and avoid scrutinizing the like-minded ones}'' [\textbf{P2}], which plays directly in the hands of the ``misinformers.'' 

As for the ``misinformers'', our participants unequivocally identified the state-sponsored ``appropriators'' that hijacked the original hacktivist playbook to spread \textit{external propaganda} on social media. That nation-states enjoyed a reputation for promulgating disinformation in the past was not a news to the hacktivists (e.g. ``\textit{Russia has always been really good at it}'' [\textbf{P2}]), but instead what caught them aback was the ``\textit{audacity and the sophistication}'' [\textbf{P4}] in utilizing trolling and memes on such a massive scale \cite{Zannettou2020}. Reflecting on this shift in online operations, \textbf{P3} believes that ``\textit{disinfo ops [sic] and hacking our intellectual property is all these nation-states are left with because they can't beat us militarily or economically}.'' Not necessarily neoliberal, but nonetheless authoritarian, the elites behind the external propaganda in equal degree conflicts with the \textit{mistrust of authority and promote decentralization} hacker postulate because is a ``\textit{blatant effort to control the social media turf and the mass of population spending their time there} [\textbf{P15}]. The external propaganda nature of disinformation also conflicts with the \textit{all information should be free} hacker postulate in the view of the hackers in our sample because ``\textit{overshadows and complicated an access to other more factual or useful information}'' [\textbf{P2}].

\section{Active Countering of Misinformation} \label{sec:results-rq2}
Literature on misinformation focuses on helping the social media \textit{users} discern falsehoods with strategies for ``pre-bunking'' i.e. forewarning and preemptive refutation of the falsehoods \cite{Lewandowsky-Inoculation} or ``debunking'' i.e. providing users verifiable corrections of the falsehoods from credible sources to break the illusion of truth \cite{Chan-debunking, Ecker-Nature}. An \textit{algorithmic} is also available for the mainstream social media platforms (the alternative ones do not deem misinformation as a problem \cite{gettr-paper}) that leverages natural language processing, image analysis, or metadata to detect trolling and memes \cite{TrollHunter, TrollHunter2020, Theisen}. Platforms also have the option for algorithmic ``soft'' moderation by either obscuring trolling and memes with warnings covers or attaching warning labels \cite{Vance-Warnings, context2022} and ``hard'' moderation for removing or suspending misinformer accounts \cite{Kumar}. None of these solutions, however fends of troll farms and meme disseminators effectively, so we wanted to know what hacktivists have to propose instead in the second research question.

\subsection{Leaking, Doxing, and Deplatforming}
Suspending user accounts by social media platforms for breach of their code of conduct is referred to as ``deplatforming'' \cite{Ali}. In the context of hacktivism, it takes a border meaning as hacktivists do investigative work that entails leaking and doxing but also confrontation with the misinformers that, in their subjective view, breaches the vision of democratic Internet. For example, hacktivists did a massive API scrapping of the alt-platform Parler to leak data that tied users to the Capitol Riots and the QAnon conspiracy \cite{Redding}, which in turn resulted in a massive account deplatforming on Twitter \cite{Bromell}. These activities spur operations to confront and expose the QAnon conspirators on social media (e.g. @QAnonAnonymous \cite{Conner}), amongst which some of our hacktivists have a direct role in ``\textit{dismantling the Qanon infrastructure}'' [\textbf{P2}]. 

The deplatforming targets political misinformation campaigns where our hacktivists ``\textit{compiled and leaked dossiers on individuals spreading hateful propaganda and those who seek to sow the seeds of violence}'' [\textbf{P1}] on social media. These operations were targeted both on ``\textit{individual spreaders, nation-states, even companies with murky records}'' [\textbf{P2}]. Several mentioned their direct operations for exposing disinformation relative to the ``\textit{Ukrainian conflict}'' [\textbf{P5}], praising the work of the Ukrainian IT Army outfit for dispelling the myth that Ukraine is committing genocide against Russians in the Donbas region \cite{Cornelius}. Hacktivists were dedicated in ``\textit{doxing companies and governmental agencies in response to the political meddling in the US internal affairs from places like Russia, Iran, and China}'' [\textbf{P8}]. Misinformation ``\textit{sanctioned by the governments}'' was targeted by hackers in attempts to deplatform prominent ``\textit{disinformation front agents on social media, like Irina Tsukerman, for example}'' [\textbf{P3}]. 

Leaks and doxing were equally utilized for misinformation beyond political counter(argumentation) and external propaganda. One of the hacktivists has dedicated considerable time on exposing cryptocurrency scammers on social media and elsewhere, deeming the feeling of it as ``\textit{better than sex}'' [\textbf{P5}]. Another was focused leaking personal details about predators on social media that spread misinformation to cover their sexual harassment and cyberstalking towards women, ``\textit{exposing both their sock puppet accounts and their real name on Twitter}'' [\textbf{P3}]. Another pushed back against criminal misinformation by doxing ``\textit{bullies, liars, and fraudsters}'' [\textbf{P20}] and one ``\textit{anti-cancel culture in case of minors}'' hacktivist noted that they ``\textit{successfully deplatformed major participants in hate campaigns and stalking of minors}''  on social media [\textbf{P5}].

\subsection{Anti-Misinformation ``Ops''}
The hacktivists in our sample engaged in misinformation saturation ops, true to the their commitment to ``fight misinformation with more information.''  One of the hacktivists stated that it is ``\textit{expected from the hacktivist community to combat misinformation in such a way}'' and noted that ``\textit{it is the sole reason they maintain a Twitter account}'' [\textbf{P3}]. Another one seconded this posture noting that ``\textit{it is frustrating to see misinformation from others and other creators but that is the main reason I continue to post on TikTok}'' [\textbf{P17}]. In the words of \textbf{P2}, ``\textit{there is more ideological aspect of it when I am fighting disinformation},'' directly invoking the mission of the true hacktivists to become reflexively ``\textit{loud and determined}'' to speak true information in response to the ``\textit{general assholery of misinformation on internet}'' [\textbf{P9}].   

Partaking in operation \#NAFO (North Atlantic Fellas Organization) dedicated to countering Russian propaganda and disinformation in Ukraine by weaponizing memes \cite{Scott}, our participants materialized a combination of saturation and doxing to ``\textit{curtail misinformers' ability to gain followers}'' [\textbf{P1}]. They extended their work to counter ``\textit{extremists and fascists and their toxic conspiracy theories}'' [\textbf{P1}] by disrupting their funding and deplatforming prominent followers, true to the spirit of the ``Antifa'' hacktivist counterculture \cite{Xu}. In a similar vein, one of the hacktivists proclaimed that they ``\textit{greatly contributed in the \#OpJane operation}'' [\textbf{P10}]. \#OpJane is the latest operation launched by Anonymous against Texas for enacting the anti-abortion Bill 8 that allows ``abortion bounty'' for anyone who will investigate and report abortion in the state of Texas \cite{Goforth}. Interestingly, in the announcement of the operation, Anonymous calls for ``fighting misinformation with enough plausible and difficult to disprove misinformation'' to make any data these bounty hunters gather as useless \cite{OperationJane-Anon}.

\section{Misinformation Evolution} \label{sec:results-rq3}
As there is virtually no cost of disseminating misinformation \cite{Mustafaraj}, it is unlikely that the online discourse will shed off the alternative narratives soon. If this gloomy prediction will eventually materialize \cite{McIntyre} or the Internet will improve because the new technologies will upgrade public's ability to judge the quality and veracity of content \cite{Anderson}, remains an open issue. Because the hacktivists are nonetheless stakeholders in resolving this issue, our third research question aimed to bring their prediction of how online spaces will fare with trolling, memes, and falsehoods in the near future.

\subsection{Counter-Misinformation Tactics}
The hacktivists in our sample unanimously posit that ``\textit{it is hard for social media platforms to keep up with removing it, so people stepping in to help is going to be of critical importance}'' [\textbf{P13}] for preserving a healthy discourse. The mobilization for ``\textit{justice and truth as a cause}'' [\textbf{P15}] is important not just for curbing misinformation but ``\textit{reclaiming information back from the political hold}'' [\textbf{P1}]. To help ``\textit{expose misinformation charlatans}'' [\textbf{P4}], hacktivists call for maintaining a code of conduct where ``\textit{no leak, doxing, or exposure action should cause anyone else harm (physical, reputation, mental)}'' [\textbf{P3}].

To begin with, \textbf{P3} recommends that we should ``\textit{stop treating disinformation as a freedom of speech}.'' As misinformers usually use this cloak to act very aggressively on social media, the next step is to ``\textit{identify what their weakness are and what triggers them - deplatforming or provocation?}'' [\textbf{P14}]. If the misinformers are unresponsive spreaders, then ``\textit{exposing, doxing, and putting their real faces through OSINT}'' [\textbf{P15}] is in due place not just on mainstream social media but also alt-platforms, forums and everywhere on Internet. If they itch for a provocation, then  ``\textit{orchestrated saturation}'' [\textbf{P5}] might work better with ``\textit{shitposts, absurd trolling, and ridiculing memes}'' [\textbf{P18}]. Here, the hacktivists note, it is vitally important to \textit{a priori} distance from a ``\textit{political whataboutery}'' [\textbf{P14}] and avoid ``\textit{coming across as censorship, disagreement, canceling that only could cause argument or dismissal}'' [\textbf{P5}]. 

Some of the hacktivists were on the opinion that ``\textit{doxing is not hacking anymore per se because you can get stuff with a credit card and documents could be easily faked nowadays}'' [\textbf{P1}]. One possible tactic, proposed by \textbf{P1}, was to ``\textit{find exploits, vulnerabilities in their platforms and step-by-step expose misinformers' amateurish way of doing trolling, using bots, and feeding think tanks to get a credibility behind their propaganda}.'' Another tactic, proposed by \textbf{P2}, was ``\textit{doxing for the purpose of having advertisers pull from supporting known misinformer influencers, like for example in the case of Andy Ngo}.'' Proposing more of a hybrid hacktivist tactics, \textbf{P4} suggested ``\textit{a latent, yet coordinated psychological warfare where psychologists rip apart these people, conduct serious OSINT to find incriminating leaks on them, and even pay for billboards and radio ads to publicly shame them}.'' Along these lines, \textbf{P11} even suggested throwing the book at them, \textit{targeting them with a social engineering attack and attempting to compromise a piece of their core infrastructure, be that their servers, Internet access, or bot credentials}.''

\subsection{Misinformation Literacy}
Hacktivists in our sample echo the sentiment regarding the social media users' susceptibility to false information found in scientific literature: laziness to check facts [\textbf{P2}] \cite{Pennycook1}, resistance to authoritative suggestions [\textbf{P7}] \cite{Kaiser}, allegiance [\textbf{P13}] \cite{Swire-Thompson}, and simple ignorance [\textbf{P16}] \cite{Cho}. As people that resort to action, hacktivists do feel the obligation to propose ways for addressing this susceptibility. In the view of \textbf{P5}, ``\textit{misinformation needs to be seen as something everyone is being watched for, and not just one group of people on the left or the right,}'' A ``misinformation social contract'' \cite{Zimdars} necessitates interventions such as ``\textit{a critical thinking curricula in schools}'' [\textbf{P18}], ``\textit{teaching hacking OpSec skills as social responsibility and rise to action}'' [\textbf{P5}], and ``\textit{forcing professional communication norms on platforms}'' [\textbf{P16}]. 

As hacktivists have little control over these interventions, they were happy to help with a development of ``\textit{truth-spreading bots for a `standoff' with misinformation-spreading bots}'' as something that could append the practice of leaks, doxing, and exposure [\textbf{P13}]. They recognized that these ``truth-spreading bots'' must help ordinary users to better find and locate facts, as information literacy is the single most effective one in dispelling falsehoods \cite{Jones-Jang}. Hacktivists reiterate that platforms do have to let ``\textit{misinformation to float on social media and make bots visible, so they gets overwhelmed with factual information}'' to demonstrate to ordinary users how to do help themselves [\textbf{P14}].

Regardless if it these stances are realistic or not, the hacktivists in our sample believe that the current approach to raising misinformation literacy is ineffective because it does not signal an ``\textit{unbiased attitude}'' [\textbf{P7}] to the social media users in the wrong. Instead of an educational and respectable tone, ``\textit{rather a `cancel culture' infused or a `your opinion is wrong' tone}'' [\textbf{P3}] plagues any attempt to help people to navigate and locate factual information. Rejection of misinformation, as a result of misinformation literacy, must come as an agreement that ``\textit{scientific facts do not have political properties, even if the social media platforms inherently do}'' [\textbf{P5}].

\subsection{Misinformation hacktivism}
The participants in our sample acknowledge that orchestrated \textit{misinformation hacktivism}, bar individual instances of ops against misinformers, is largely absent from social media. For the hacktivists to assume misinformation as a worthy cause for action, the conflict between the past ``hacking for political causes'' and \cite{Karagiannopoulos} future ``hacking against using falsehoods in furthering political causes'' \cite{Conner} must be resolved. Though this conflict is complex and evolving, several of the hacktivists worried that it could nevertheless create a ``\textit{division between the hacktivists on political lines}'' [\textbf{P2}]. 

As a relative threat to the misinformation activism, one participant mentioned the hijacking of the hacktivists image for self-promotion, e.g. ``\textit{some like to portrait themselves as woke gods of the web with zero fuck-ups}'' [\textbf{P12}]. Another threat is the temptation of using misinformation against misinformation, as in the \#OpJane campaign [\textbf{P10}]. While this strategy is true to the ``fight-fire-with-fire'' approach, it might backfire in circumstances where abiding to the hacktivist ethic comes secondary to expressing social and political angst on social media \cite{Mihailidis}. On top of this, one could argue that this conflict \textit{per se} might be hard to resolve in the misinformation instance as external propaganda, because even if the hacktivists are ``hacking for the homeland,'' they nonetheless are doing it on political terms \cite{Dahan}.

\section{Discussion} \label{sec:discussion}
\subsection{Implications}
The new brand of misinformation, our findings show, draws the ire of the hacktivists, reprehending the hijacked discourse for political and propagandistic proposes. The ``fight-fire-with-fire'' response -- leaks, doxing, and deplatforming -- though individually employed by some of the participants in our sample, is yet to be orchestrated and tested against serious disinformation outfits that, unfortunately, are still out there on social media \cite{Im}. The early evidence outside of the US shows that this orchestration works as the IT Army leaked data from Russian organizations in response to the troll farms' disinformation narrative that Ukraine is committing genocide against Russians in the Donbas region \cite{Cornelius}. 

The hacktivists' resoluteness to go after the misinformers would certainly have implications for the content/user moderation on social media, user participation, and future of Internet activism overall. Moderating users and content on social media was, and still is, the response by the mainstream platforms to the political and public health misinformation \cite{context2022}. Alternative platforms like Gab, Gettr, and Parler, seen as the seeding grounds for this misinformation \cite{Blackburn}, on the other hand, never did, nor currently do, employ any content/user moderation \cite{gettr-paper}. While the content/user moderation incites a migration from the mainstream to the alt-platforms \cite{Blackburn}, it remains to be seen whether the deplatforming will have the same effect. Mainstream social media had a mixed response to leaks and doxing in the past (e.g. allowing WikiLeaks \cite{Sifry} and barring the Hunter Biden's laptop leaks \cite{Diesen}), so this also adds uncertainty if and how the hacktivists' ``fight-fire-with-fire'' approach will be allowed, moderated, or perhaps even forced to migrate entirely outside of the social media space.

Trolling and memes might still maintain the popularity amongst the misinformers, however, the latest modes of social media participation like short videos on TikTok open new ``fronts'' for both the misinformers and the hacktivists. TikTok has increasingly been tested as the next ``battlefield'' of alternative narratives with evidence of health and abortion misinformation \cite{TikTok-abortion, Basch} and an individual engagement by at least one of participants. Recalling that the hacktivists' \#OpJane was waged in response to the abortion ban laws in Texas and called for ``misinformation-against-(mis)information'' \cite{Goforth}, it is yet to see how the leaks, doxing, and deplatforming will materialize with meme-ified videos and trolling. TikTok claims it does health and abortion misinformation moderation \cite{TikTok-safety}, but evidence shows that this is lax and largely ineffective \cite{newsguard}, adding an additional incentive for shifting the disinformation campaigns on this platform. 

TikTok is also the next platform for Internet activism where the hashtag activism is appended with videos expanding the developing news narratives, such as the coverage of the Black Lives Matter movement and the Capitol riot \cite{Literat}. TikTok presents content not just from viral hashtags but also their variations (e.g. \#abotion but also \#ab\^{o}rtion \cite{TikTok-abortion}) so the threat of hashtag hijacking, co-opting, and counter hash tagging will inevitably materialize here too. This particular affordance likely will allow for weaponizing deepfakes in appending the hashtag war in near future, as they already appeared in misinformation videos about the COVID-19 pandemic on TikTok \cite{Shang}. All of these developments would certainly necessitate a dynamic adaptation in the way doxing, leaking, and deplatforming is performed in order not just to avoid disintegration of the Internet activism and hacktivism, but prevent another paucity in action that brought the state-sponsored misinformation \textit{en masse} on social media in the first place \cite{Hannan}.

\subsection{Ethical Considerations}
The purpose of our study was not to generalize to a population; rather, to explore the contemporary hacktivists' relationship with misinformation in depth. To avoid misleading readers, we did not report percentages, names, or tools, tactics, and procedures mentioned during the interviews. A full moral evaluation of the suggested countering and/or utilizing misinformation is out of scope of this paper, though we condemn any action of leaks, doxing, exposure, or rumors that could result in an individual harm of any form. We are careful with our study not to infringe upon the hacktivist's aesthetics nor to cause any negative actions with our findings. 

We would, however, point out that our engagement with, rather than a disavowal of, the hacktivists can help in refining and revisiting some of the over-simplistic hacktivism portraits of toxic vigilantism, nihilism, and criminality \cite{Goode}. While we maintain that each operation -- misinformation hacktivism related or not -- has to be morally justified separately, we find reasonable to identify with the ideas and suggestions put forth by the hacktivists in our study, as they are in conformity with the Levy's hacker ethos \cite{Levy} and the democratic vision of the Internet \cite{Hatta}. We also accept and support the idea of ``fight fire with fire'' action identified in our findings, as it seeks to fill a resistance void arising from a scale mismatch between institutional regulation, lax participation policies and perverse incentives of all the platforms, as well as the experience of living with misinformation in our everyday discourse \cite{Inglott}.

\subsection{Limitations}
Our research was limited in its scope to U.S. hacktivists, therefore we exercise caution to the generalization of the results across the entire Internet activist community worldwide. Many hacktivist operations are often in the center of debates regarding the dimensions of civil disobedience, political participation, legality, and the ethical use of Internet technologies \cite{Inglott}. Our results pertain neither to append the permissiveness nor the disapproval of the these operations, rather, to voice the opinion of the hacktivists as the unique and engaged Internet minority. Even with such a relatively small sample we gathered in our study, we got a wide variety of insights to which many other hacktivists could well disagree and propose other models, approaches, and visions of dealing with misinformation. 

We are aware that our results represent the contextualization informed by all forms of misinformation that currently exist on social media. Therefore, we are careful to avoid any predictive use of our results in future misinformation campaigns. Importantly, we do not know if, when, and how the hacktivists in our sample used the proposed counter-misinformation tools, tactics, and procedures. Our results do not provide blanket justification for any frivolous use of them across social media and any other online spaces. We note that this study reported on the evolving experience of dealing with misinformation by hacktivists and might miss some important aspects of meting out the truth on social media. We advise caution to this, as we see our work as a synergistic line of scientific inquiry addresses an important gap in voicing the opinions of those that actually introduced the means for mass producing of misinformation online in the first place. 

\subsection{Future Work}
Our future research will continue to trace the way the hacktivist community engages with misinformation. We are interested to expand our work beyond U.S. and work with hacktivists across the globe, as misinformation is contextual to the geopolitical makings in the space where many of them operate. We are set to further explore the intersection between hashtag activism and hacktivism for the same cause of countering misinformation as such synergistic activities do already emerge in some form, as the case with the \#NAFO campaign on Twitter. Here, we would devote much attention to the new misinformation ``battlefield'' of platforms for short videos such as TikTok and Instagram. It would be useful to study the emergent circumstances in which misinformation hacktivism mobilizes and empowers ordinary users to join future ``Troll [target] Day'' operations and catalogue their experiences with such participation. Of equal importance, too, would be to further study the use of ``misinformation-against-(mis)information'' as in the case of \#OpJane to learn both the benevolent and potentially malevolent aspects of this approach. 

\section{Conclusion} \label{sec:conclusion}
Reflecting the communitarian ideals of free information and disobedience to authority, the hacktivists in our study showed a determination for a radical response against the reprehensible act of spreading falsehoods on social media. As misinformation is consequential to the trolling and memes of the early days of hacktivism, it is appreciative to learn that the contemporary hacktivists are outwardly against such a nefarious appropriation of their aesthetics. It is encouraging to reveal that hacktivists also advocate for general misinformation literacy as a strategic asset against an undemocratic Internet. These findings, we hope, will empower ordinary users who share the same action space in reprobating misinformers for the sake of maintaining the vision of democratic Internet.

\bibliographystyle{plain}
\bibliography{\jobname}

\newpage
\section*{Appendix} \label{appendix}
\begin{enumerate}
\itemsep 0.5em
    \item How do you describe your niche, role, activity, or agenda you have online?
    \item What brought you to hacking, OSINT, cyber-threat intelligence, and any operations you have taken so far?
    \item Have you faced any obstacles, challenges, repercussions because of your activity?
    \item Has the obstacles, challenges, repercussions affected your commitment, motivation, and vision of your actions and in what way?
    \item What is your take on the increased misinformation proliferation online?
    \item Have you ever engaged or considered engaging in utilizing your actions in exposing disinformation campaigns? What was the disinformation about, in what capacity you participated, and what were the outcomes you were attempting to achieve?
    \item What do you think the tools, tactics, and procedures undertaken in a hypothetical \textit{misinformation hacktivism} operation might entail?
    \item What in your opinion, is the way to continue evolving this work and in what shape and form?
    \item Is there anything else that you would like to add or say that is relevant to the questions we have asked so far?
    \item If you would like to share some demographic information, please do - we don't require it but it will help us better contextualize your effort and story. 
\end{enumerate}

\end{document}